\documentclass{article}
\def\blankline{\vskip 12 pt\noindent}
\usepackage{lineno}
\usepackage{graphicx} % Required for inserting images
\title{Cosmic-ray characterization of a timing detector based on extruded scintillators and wavelength-shifting fibers read-out through SiPMs}
\author{F.Anulli, A.Betti, C.Bini, M.Corradi, C.Luci, L.Martinelli, A.Nisati,\\ M.Pantalena, S.Rosati, R.Vari, S.Veneziano \\ \\ Sapienza Universit\`a di Roma and INFN Roma}
\begin{document}
%\author[]{}
\maketitle
\begin{abstract}
     Timing detectors based on extruded scintillators of different dimensions with different kinds of wavelength-shifting fibers read-out through SiPMs have been extensively tested using a cosmic ray telescope. Results in terms of light-yield, time resolutions and longitudinal coordinate resolutions are presented.
\end{abstract}
\section{Introduction}
\par\noindent The proposed FCC-ee electron-positron facility at CERN \cite{FCCee} has been identified by the international scientific community
as the future particle collider in Europe after LHC and its upgrade in luminosity, HL-LHC.
The precise detection and measurement of muons is an important component of the physics program at FCC-ee. Muons provide a powerful tool for high precision measurements and test of the Standard Model, in particular of the electroweak and Higgs sector, as well as the flavor sector.
High-performance muon reconstruction is also essential both for maximizing the sensitivity to potential signatures of physics
beyond the Standard Model, including rare processes, indirect effects of heavy new particles, and exotic processes \cite{briefing}.
\par\noindent Different approaches for the measurement of muons are considered in the different experimental proposals under definition. In one of these proposals \cite{Allegro}, precision drift tubes are combined with fast plastic scintillators to enable both spatial and timing measurements. These scintillators should provide a time resolution of the order of a few hundreds of ps for triggering and time of flight measurements and a space resolution of the order of few mm in one direction to provide second coordinate measurement, and of few cm in the other direction to reduce possible ambiguities in tracking. 
\blankline
\par\noindent In the last decades, the detection technique based on plastic scintillators has considerably improved: a novel kind of photon detectors, the SiPM became the standard in photon detection; “low-cost” extruded scintillators have been produced in large quantities with different shapes and dimensions to be used in conjunction with wavelength shifting fibres (WLS) in several projects; the technology of WLS fibres evolved in the direction of long attenuation lengths (above 3-4 m) and short decay times (down to the ns level). Several detectors have been built with this concept \cite{Pyramid, Belle2, Russi1, Russi2, Mu2e, Bross}.
\par\noindent Based on that, scintillator-based detectors can now be considered to be competitive in cost and in robustness other than in time resolutions and in rate capability, for large surface muon detectors at next generation colliders. Such detectors should be operated in conjunction with high spatial precision detectors providing a good measurement of the second coordinate and a high precision Time of Flight measurement.  
\blankline
\par\noindent This paper presents a study of the performance of extruded scintillators coupled to different types of WLS fibers read-out by SiPMs. The study ws performed using a cosmic-ray telescope. The focus is on the light-yields and time resolutions that can be obtained in different detector configurations. The selected configurations span the range of scintillator thicknesses and WLS fiber characteristics currently considered realistic for large-area muon systems
\section{Experimental set-up}
\par
\begin{figure}
    \centering
    \includegraphics[width=0.8\linewidth]{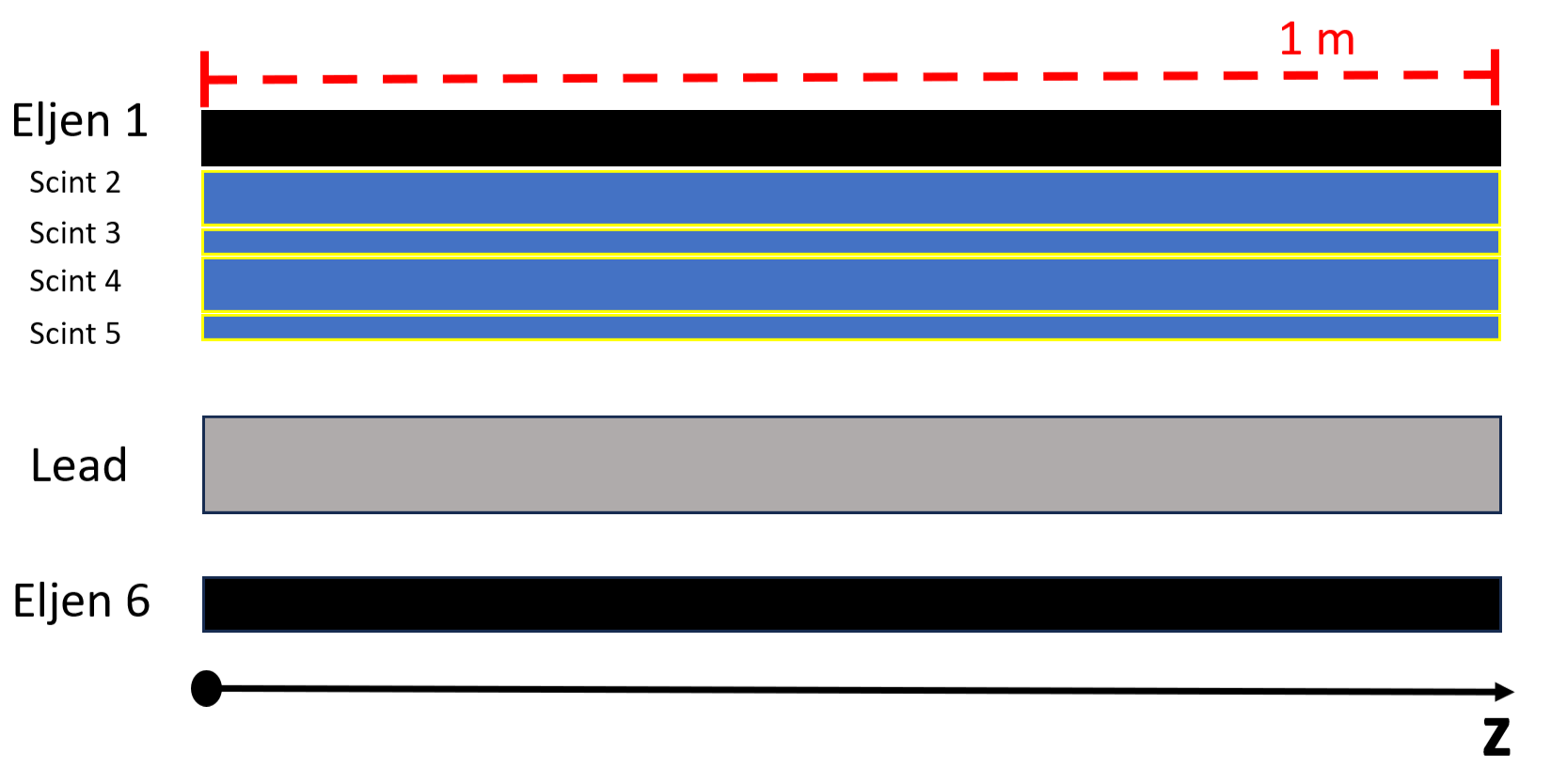}
    \caption{Schematic view of the experimental set-up. The two Eljen bars, 1 and 6, provide the trigger signal; Scint 2 to 5 are the four bars under test; the Lead absorber allows to select high momentum cosmic muons. The $Z$ coordinate, running along the fiber direction is defined.}
    \label{lay-out}
\end{figure}

\noindent The tests were performed using a cosmic-ray telescope at the INFN laboratories in Sapienza University in Rome. The telescope, schematically shown in Fig.\ref{lay-out}, is based on two 1 m long bulk scintillation counters EJ-200 10x20 mm$^2$ cross-section area providing a trigger. The two counters are read-out on both sides by Hamamatsu SiPMs S136360-6075PE of 6x6 mm$^2$ sensitive area directly facing the two sides. The trigger signal requires the coincidence of the four signals, each of them exceeding a threshold of 30 mV on a time window of about 200 ns. Between the two trigger counters, a layer of 5 cm lead blocks is inserted, just above the lower counter. The extruded scintillators under test (extruded bars in the following) are placed between the upper counter and the lead absorber. The entire structure is closed in a light-tight box.
\par\noindent The signals from the two trigger counters are read-out through a DRS4 module \cite{DRS} and are used to select cosmic ray tracks that are almost vertical in both directions. Typical trigger rates are of 0.2 - 0.3 Hz.
\blankline
\par\noindent Two kinds of extruded bars of different sizes have been tested, all provided by the FNAL-NICADD Extrusion Line Facility at Fermilab \cite{Fermilab}: 
\par\noindent "Thick" bars 1 m long with a rectangular cross-section of 16x19 mm$^2$ and a 2 mm diameter hole placed at the rectangle center; 
\par\noindent "Thin" bars also 1 m long but with a rectangular cross-section of 6.5x25 mm$^2$ and a 1.5 mm diameter hole at the rectangle center. 
\par\noindent All bars use a co-extruded titanium reflective coating to maximize light collection. 
\par\noindent Three different kinds of fibers are used, namely YS-2, YS-4 and YS-6, all manifactured by Kuraray \cite{japanese}, characterized by different decay times and slightly different absorption and emission peaks, all well matching the sensitive peak of the used SiPM. The fibers are available in different diameters ranging from 1 to 2 mm. 
\blankline

\begin{table}
    \centering
    \begin{tabular}{|c|cc|cccc|}
    \hline
        Conf & Extr & Thickness & WLS fiber & Diameter & $\tau$ & $\lambda_{emis}$\\
        & & (mm) & & (mm) & (ns) & (nm) \\
        \hline
        1 & Thick & 16 & YS-2 & 2.0 & 3.2 & 474\\
        2 & Thick & 16 & YS-2 & 1.5 & 3.2 & 474\\
        3 & Thick & 16 & YS-4 & 1.5 & 1.4 & 470\\
        4 & Thick & 16 & YS-6 & 1.0 & 1.3 & 462\\
        5 & Thin & 6.5 & YS-2 & 1.5 & 3.2 & 474\\
        6 & Thin & 6.5 & YS-4 & 1.5 & 1.4 & 470\\
        7 & Thin & 6.5 & YS-6 & 1.0 & 1.3 & 462\\
        \hline
    \end{tabular}
    \caption{The configurations under test are numbered here from 1 to 7. For each configuration we give: the type of extruded scintillator with its thickness and the type of WLS fiber with its diameter, decay constant and emission peak wavelength.}
    \label{configurations}
\end{table}
\par\noindent The configurations that have been tested are defined and listed in Table \ref{configurations}. The properties of the corresponding fiber are also shown.
\begin{figure}
    \centering
    \includegraphics[width=0.7\linewidth]{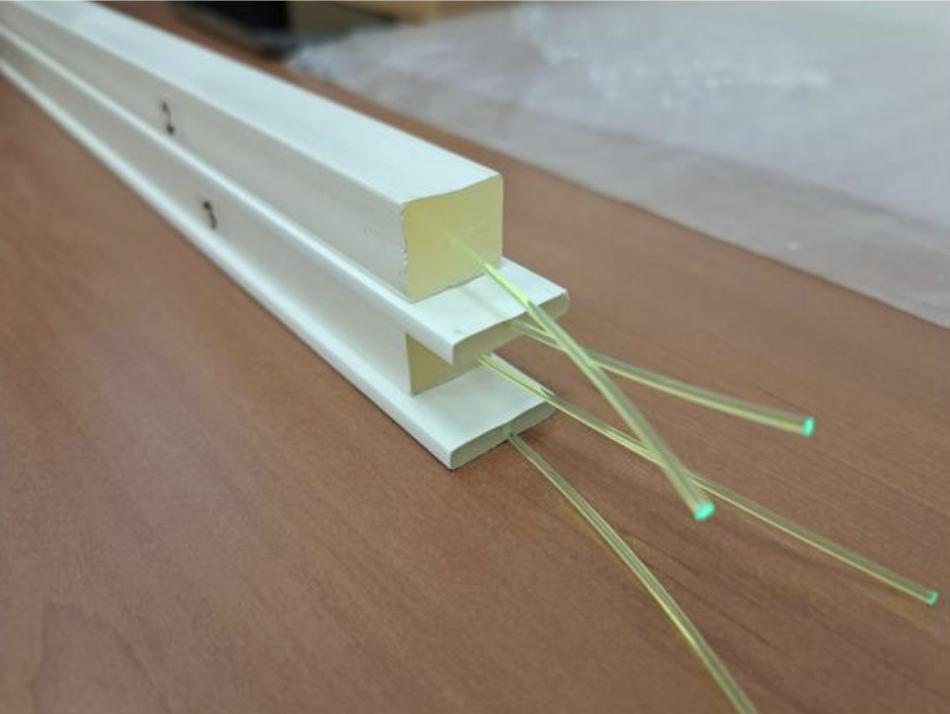}
    \includegraphics[width=0.7\linewidth]{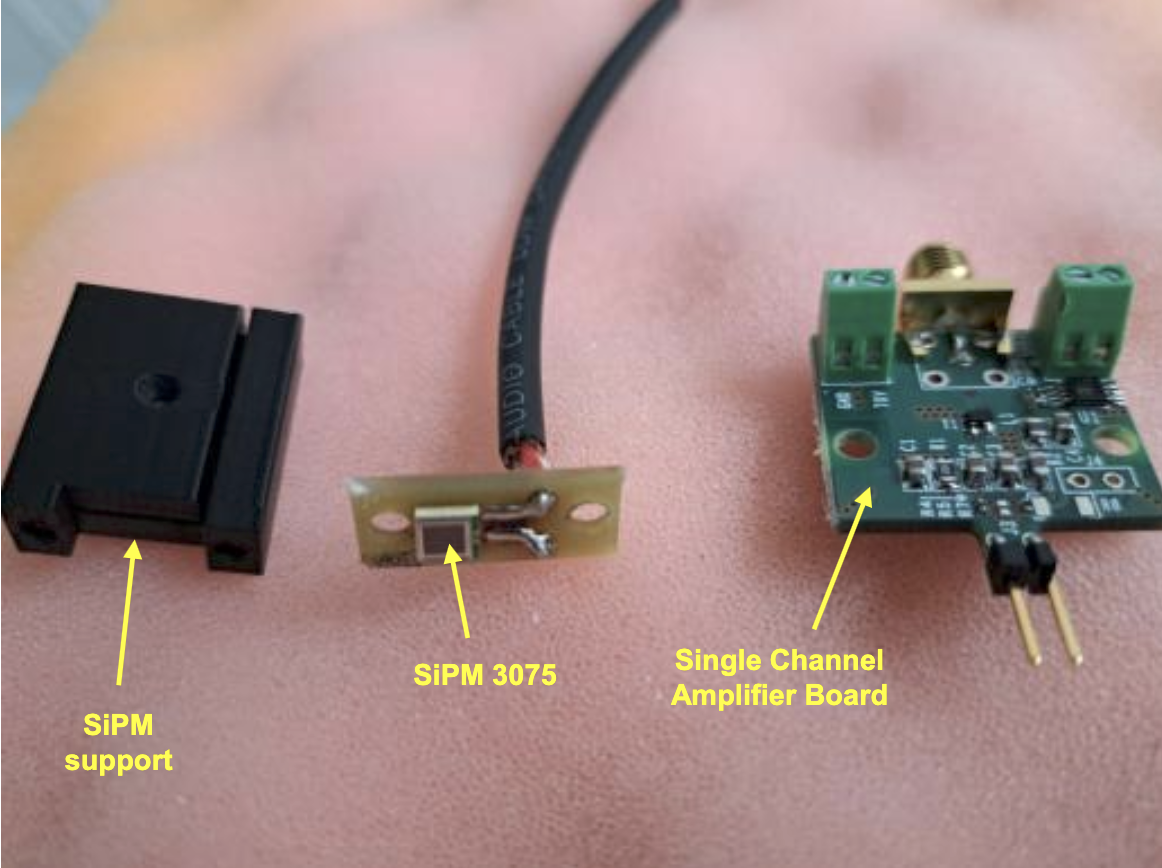}
    \caption{(Upper) Example of four bars under test. It can be seen the bar shapes and the fibers inserted in the holes. (Lower) From left to right: SiPM support to guarantee fiber-SiPM matching; SiPM base with wire for HV bias and for signal read-out; a single channel Amplified board connected to the SiPM.}
    \label{strips}
\end{figure}
\par\noindent A picture illustrating the extruded bars under test is shown in Fig.\ref{strips}. Once cut using a simple cutter, the fibers are polished at each end using first a \#2000 sandpaper with water and then a flannel cloth layer. The fibers are directly inserted into the hole without any material around them, even when the hole diameter exceeds the diameter of the fiber. Each fiber end is in direct contact with the surface of the SiPM. A test using optical grease between the fiber end and the SiPM surface has been done and the result is shown in the following. However, in the data presented in this paper no optical grease is used so that the contact is guaranteed by fixing the position of the fiber with respect to the SiPM surface through a specifically designed support also shown in Fig.\ref{strips}.

\section{SiPM read-out and calibration}
\par\noindent SiPMs Hamamatsu S136360-3075PE of 3x3 mm$^2$ sensitive area are used to detect the light coming from the fibers. The bias voltage is provided to all SiPMs through a single channel power supply A7585D from CAEN. The breakdown voltages of the used SiPMs, provided by Hamamatsu, range between 51.9 and 52.4 V. Data were collected powering the SiPMs with 54, 55, 56 and 57 V, with HV=55 V being the baseline powering voltage corresponding in average to $\sim V_{break}$+3 V. The test is done at room temperature, ranging between 20 and 23 degrees, and no temperature correction is applied.
\par
\begin{figure}
    \centering
    \includegraphics[width=0.75\linewidth]{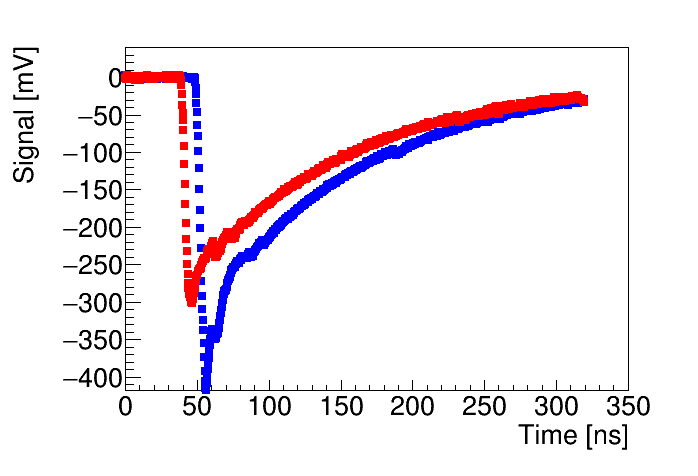}
    \caption{Two examples of signals of cosmic rays on extruded bars taken through the entire read-out chain with the CAEN Module. }
    \label{signals}
\end{figure}

\noindent Each SiPM signal is amplified by a discrete components custom design high-frequency amplifier for SiPM applications shown in Fig.\ref{strips}, and is read-out through a DT5743 8-channel 3.2 GS/s CAEN digitizer. The full shape of the signals are saved for all events in a time window 320 ns, so that time and amplitude of each signal are evaluated offline. Typical examples of cosmic ray signals are shown in Fig.\ref{signals}. The pedestal is evaluated as the average baseline voltage before the signal window, in a time range between 5 and 25 ns; the noise level is estimated as the RMS of the baseline voltage in the same range. The pedestal is subtracted event by event in the amplitude evaluation. The signal amplitude is given by the signal integral in the time range from 25 to 320 ns.
\blankline
\par
\begin{figure}
    \centering
    \includegraphics[width=0.8\linewidth]{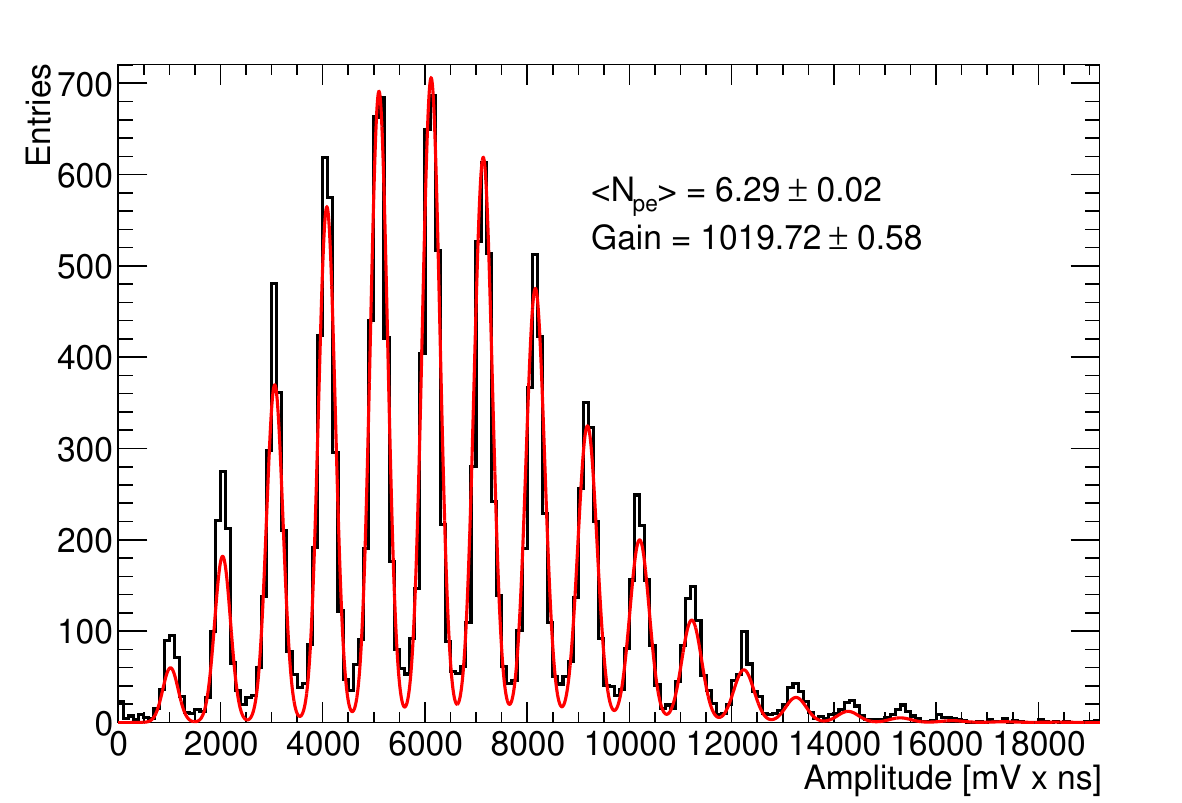}
    \caption{Example of amplitude distribution for a SiPM illuminated by a LED source. From the superimposed fit the SiPM gain can be extracted as the spacing between peaks.}
    \label{photoelectrons}
\end{figure}

\begin{figure}
    \centering
    \includegraphics[width=0.8\linewidth]{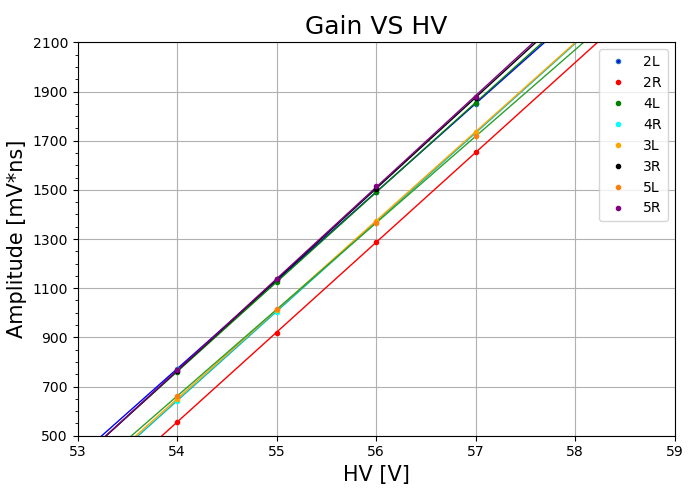}
    \includegraphics[width=0.8\linewidth]{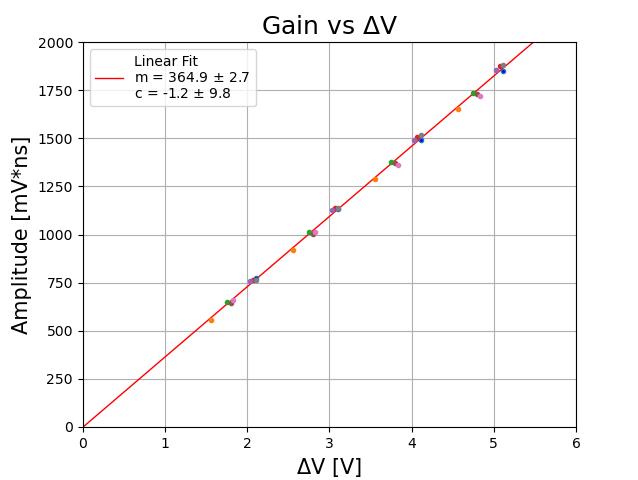}
    \caption{(Upper) Signal amplitude vs HV for the eight SiPM used for the four bars under test. (Lower) The same amplitudes are plotted as a function of $\Delta V=HV-V_{break}$, with $V_{break}$ given by Hamamatsu data sheets. Labels 2L, 2R, ..., 5L and 5R refer to the scintillator naming convention shown in the sketch of Fig. 1, where scintillator 2 is the uppermost and scintillator 5 the lowermost. The letters L and R denote the left and right SiPM readout sides, respectively.}
    \label{calibHV}
\end{figure}

\noindent In order to express a cosmic ray signal in number of photoelectrons, calibrations runs are performed using a white LED source, the photons being captured and re-emitted by the fibers. These runs allow to evaluate the amplitude corresponding to a single photoelectron (we call it "gain" in the following). Fig.\ref{photoelectrons} shows a typical LED photon spectrum for a SiPM powered at the baseline voltage of 55 V, with a fit superimposed. The fit allows to extract the gain of the SiPM as the distance between peaks. By evaluating the equality of the spacing between the peaks up to 20-25 photoelectrons we verified that using the signal integral as a measure of the signal amplitude ensures a linear dependence of the amplitude on the number of photoelectrons.
Fig.\ref{calibHV} presents the gain as a function of the bias voltage for the eight SiPMs used for the test. By subtracting the $V_{break}$ values provided by Hamamatsu, the eight curves become superimposed, as expected. 
\par
\begin{figure}
    \centering
    \includegraphics[width=0.75\linewidth]{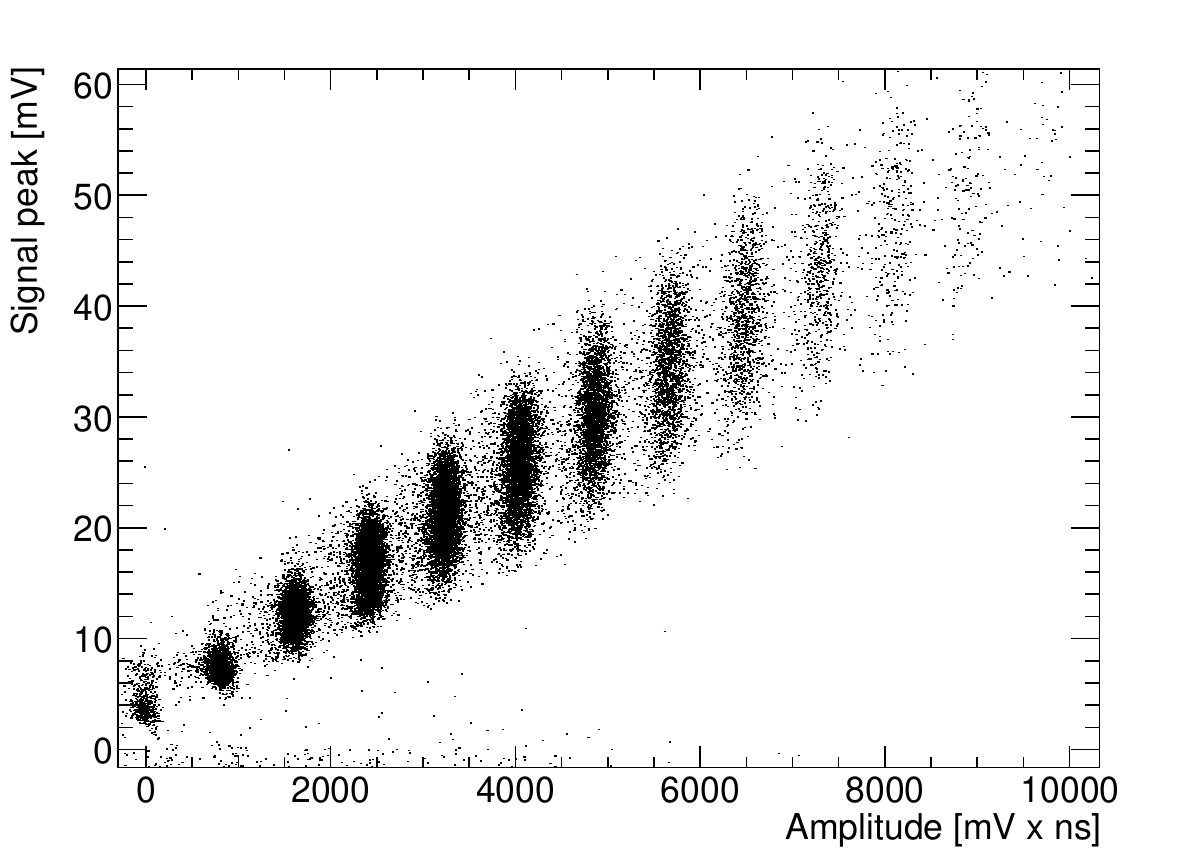}
    \caption{For a calibration run taken with LEDs at the nominal HV of 55 V, scatter-plot of the signal peak vs. the amplitude of the signal. From left to right: the pedestal spot (0 amplitude) and the spots corresponding to the different numbers of photoelectrons from 1 to 10-11. From this plot one can obtain the signal peak corresponding to a single photoelectrons, around 8 mV in this case.}
    \label{calibThr}
\end{figure}

\noindent In order to express the applied thresholds also in terms of number of photoelectrons, the signal peak for one photoelectron has to be evaluated. Fig.\ref{calibThr} illustrates, for one SiPM at 55 V, the signal peak vs. the signal amplitude. The plot shows the pedestal (first region on the left) and the regions corresponding to the different photoelectron peaks. It can be seen that the first photoelectron corresponds to a signal peak of approximately 8 mV. The observed spread of the peak distributions for increasing number of photoelectrons, is due to the spread in the arrival times of the photons. 

\section{Method of measurement}
\par\noindent In a given run, a pair of Thick bars and a pair of Thin bars are inserted between the trigger counters in alternate positions. The bars of each pair are in the same configuration, one of those listed in Table.\ref{configurations}. All bars are read from both sides and all SiPMs are held at the same bias voltage. The trigger counters are also read-out, so that in total, twelve signals are acquired in each event.
\par\noindent Typical noise values are at the 1 mV level. Signals with a noise level greater than 5 mV are discarded. Since the cosmic signals pass the threshold in a typical window between 30 and 70 ns, signals with a time outside this window are also discarded.
\par\noindent The following quantities are measured for each bar $i$: the left and right amplitudes $A^i_{L,R}$; the left and right times $t^i_{L,R}$ given by the times at which the threshold is passed. 
\par\noindent Light-yields in terms of number of photoelectrons are obtained by evaluating the peaks of the amplitude distributions as shown in Fig.\ref{peakexamples} and normalizing them to the corresponding SiPM gain as obtained in the calibration runs.
\par\noindent The bar times $T^i$ and the bar longitudinal coordinates $Z^i$ are obtained according to the relations:
\begin{figure}
    \centering
    \includegraphics[width=0.9\linewidth]{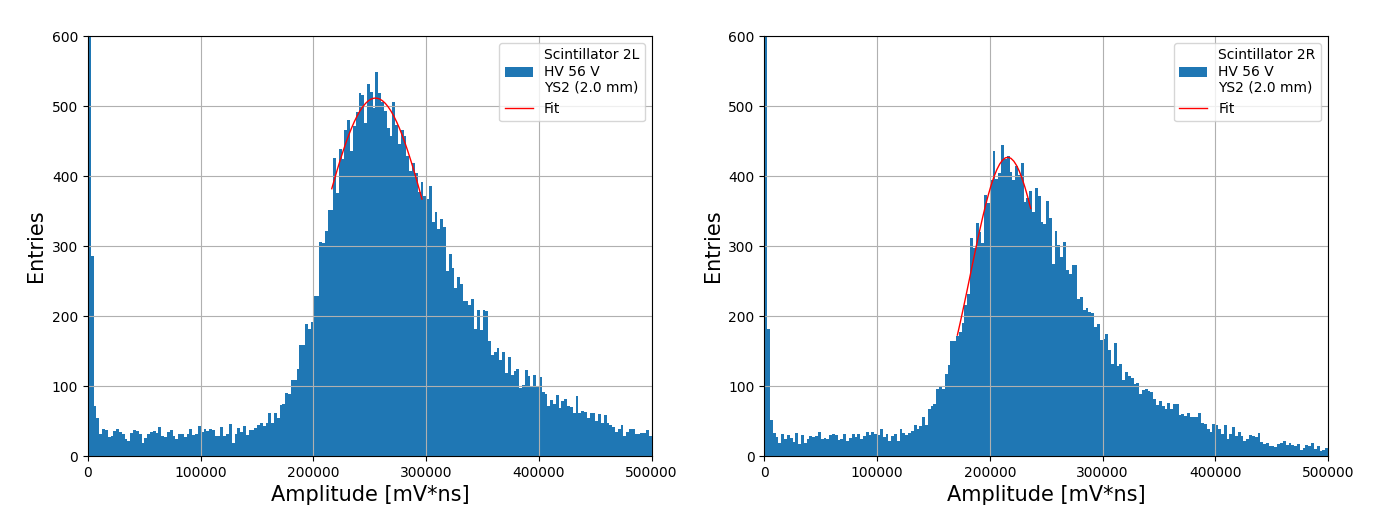}
    \caption{Examples of amplitude spectra measured by two SiPMs. The peak of the distribution is evaluated with a Gaussian fit around it.}
    \label{peakexamples}
\end{figure}
\begin{eqnarray}
    T^i=\frac{1}{2}\Big(t^i_L+t^i_R\Big)\label{Tsum}\\
    Z^i=\frac{v_l}{2}\Big(t^i_L-t^i_R\Big)
\end{eqnarray}
\par\noindent where $v_l$ is the effective light propagation speed in the fibers. The definition of the bar time $T^i$ given above allows to have a time value that is independent from the position of the particle along the bar. $v_l$ is obtained measuring the width of the $Z^i$ spectra, as shown in the next section. Values of $Z$ are evaluated with the same method for the two trigger counters, and events are discarded if the $Z$ difference between upper and lower trigger counters exceeds 6 cm.
\par
\begin{figure}
    \centering
    \includegraphics[width=0.9\linewidth]{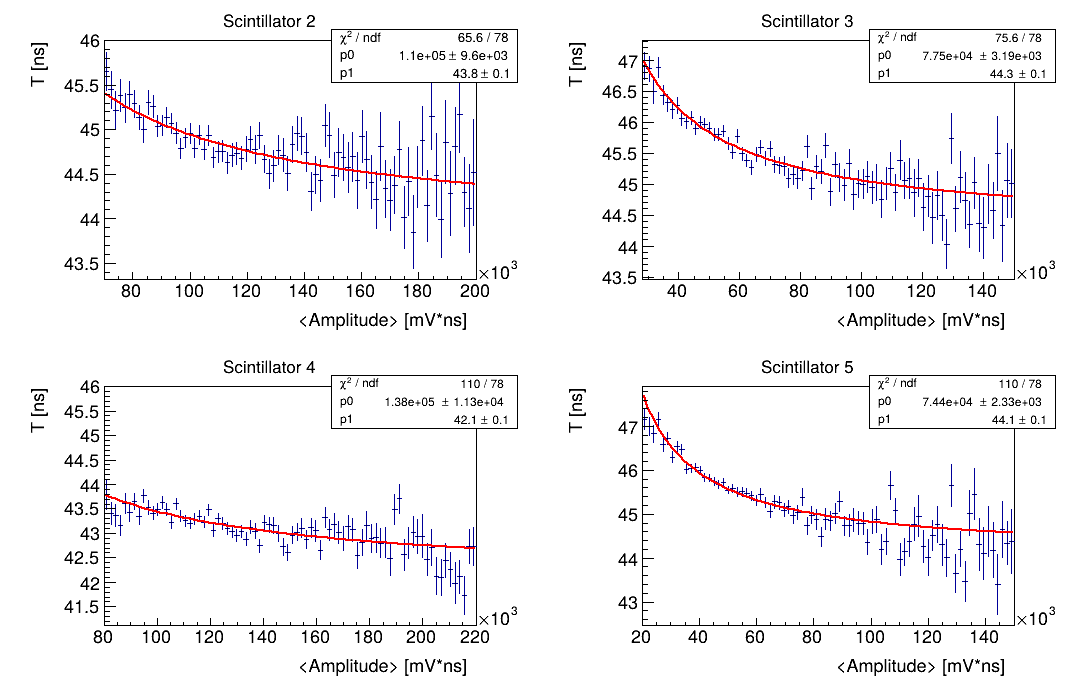}
    \caption{Dependence of the bar time $T^i$ on the average amplitude of the sides of the bar, for four different bars. The plots on the left correspond to Thick scintillators, the ones on the right to Thin scintillators. The data are fit with a two-parameter fit $p_0/x+p_1$.}
    \label{slewing}
\end{figure}

\noindent The bar times $T^i$ are corrected applying a signal slewing correction algorithm. The behavior of $T^i$ as a function of the average signal amplitude is well described by a two-parameter function $a/x+b$. Fig.\ref{slewing} illustrates few examples of such a dependence with the fit for the slewing corrections superimposed. The effect turns out to be relevant for the Thin bars and less relevant for the Thick bars.

\par\noindent Assuming that the bars in each pair are "identical", we get the $T$ and $Z$ resolutions $\sigma(T)$ and $\sigma(Z)$ respectively, by the guassian width of the distribution of the differences between the corresponding values of each pair. In this way any possible trigger jitter is canceled.
\begin{eqnarray}
    \sigma(T)=\frac{1}{\sqrt{2}}\sigma(T^{up}-T^{down})\\
    \sigma(Z)=\frac{1}{\sqrt{2}}\sigma(Z^{up}-Z^{down})
\end{eqnarray}
\par\noindent $up$ and $down$ indicate the two bars of the pair.
\par
\begin{figure}
    \centering
    \includegraphics[width=0.8\linewidth]{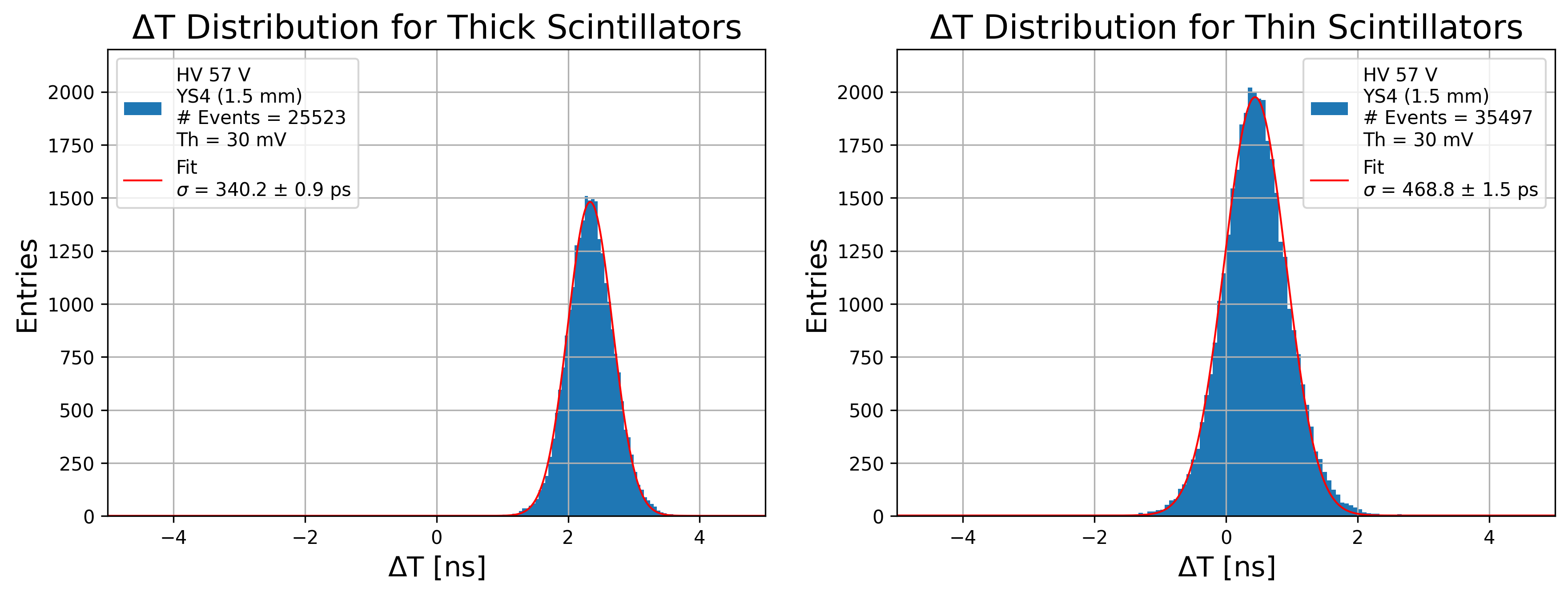}
    \includegraphics[width=0.8\linewidth]{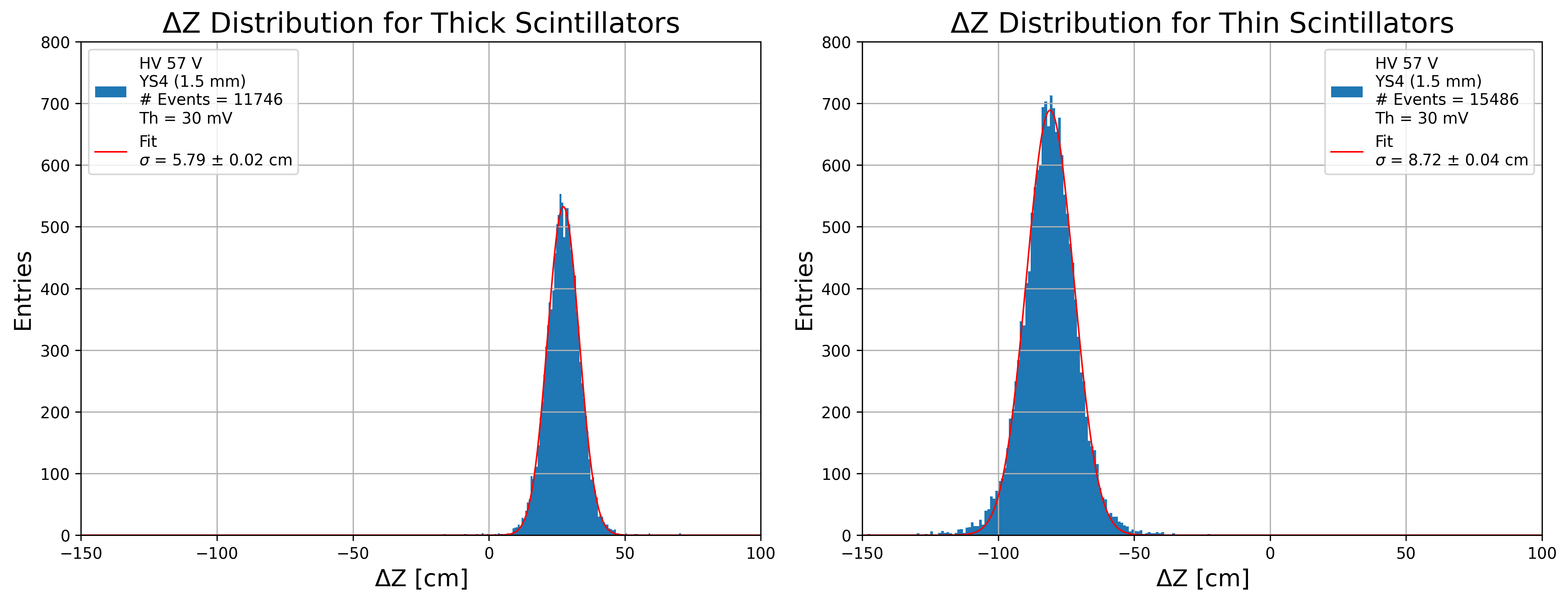}
    \caption{Examples of $\Delta T=T^{up}-T^{down}$ (upper plots) and $\Delta Z=Z^{up}-Z^{down}$ (lower plots) distributions, for Thick and Thin scintillators. In all cases single-Gaussian fits well reproduce the distributions.}
    \label{differences}
\end{figure}

\noindent As shown in Fig.\ref{differences} the typical distributions from which the resolutions are extracted are well described by Gaussian functions. All the $\sigma$s quoted in the following the paper are directly extracted from single-Gaussian fits.

\section{Data analysis and results}
\subsection{Light-Yields} 
\begin{figure}
    \centering
    \includegraphics[width=0.8\linewidth]{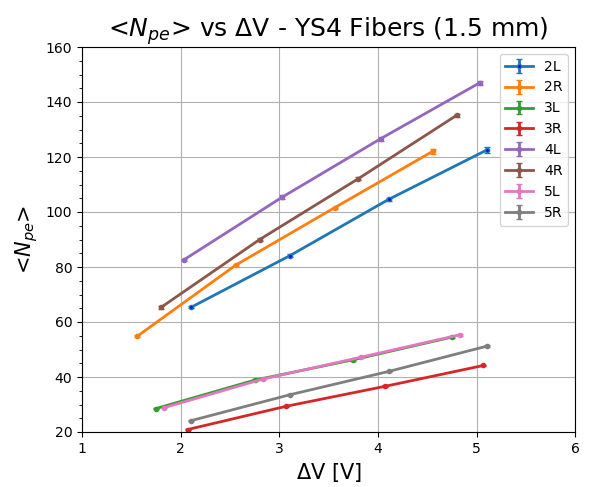}
    \caption{Light-yield in number of photoelectrons as a function of $\Delta V=HV-V_{break}$ for Thick and Thin bars in configurations 3 and 6. These curves should follow the PDE behavior of the SiPMs. The labels 2L, 2R, ..., 5L and 5R refer to the scintillator naming convention shown in the sketch of Fig. 1.}
    \label{lyexamples}
\end{figure}

For each configuration the light-yield is measured as a function of HV using the amplitudes measured by the four SiPMs corresponding to the given configuration. The behavior of the light-yield vs. $\Delta V = HV-V_{break}$ is shown as an example for two different configurations (3 and 6 of Table.\ref{configurations}) taken in the same run in Fig.\ref{lyexamples}. The light-yield increases with HV due to the known dependence of the SiPM photon detection efficiency (PDE) on HV. As can be seen, there is some spread in the light-yield of the four SiPMs connected to the two bars corresponding to the same configuration. This can be attributed to different fiber quality or fiber cut quality or by the SiPM-fiber matching. The average light-yield $<N_{pe}>$ is obtained as the average of the four SiPMs at $\Delta V=3$ V. Table \ref{lysummary} summarizes the average light-yield for each configuration evaluated at an HV=V(breakdown)+3V. 
\par\noindent From this table the following conclusions can be drawn.
    \begin{enumerate}
    \item By comparing the light-yields of Thick and Thin scintillators obtained with the same fibers we get the following ratios: 2.66$\pm$0.19 (YS-2); 2.34$\pm$0.24 (YS-4); 2.63$\pm$0.21 (YS-6), in agreement with the thickness ratio of the two scintillators 16/6.5 =2.5.
    \item By comparing the light-yields of the same fibers with a different diameter in the same scintillator we get the following ratio: YS2 (2mm)/YS2 (1.5 mm) =1.34$\pm$0.13 in agreement with the ratio of the diameters 2/1.5=1.33.
    \item Finally, by comparing the light-yields of different fibers, YS-2 and YS-4 with the same diameter, 1.5 mm, in the same scintillator we get the following ratios: YS-2/YS-4 = 1.16$\pm$0.08 (thick scintillators); 1.14$\pm$0.08 (thin scintillators). This indicates a $\sim 15\%$ larger light-yield of YS-2 fibers with respect to YS-4 fibers.
    \end{enumerate}
\begin{table}
    \centering
    \begin{tabular}{|c|ccc|c|}
\hline
        Conf & Thickness & WLS fiber & Diameter & $<N_{pe}>$\\
         & (mm) & & (mm) &  \\
        \hline
        1 & 16 & YS-2 & 2.0 & 145$\pm$12\\
        2 & 16 & YS-2 & 1.5 & 107$\pm$5\\
        3 & 16 & YS-4 & 1.5 & 92.2$\pm$4.6\\
        4 & 16 & YS-6 & 1.0 & 40.1$\pm$4.1\\
        5 & 6.5 & YS-2 & 1.5 & 40.1$\pm$1.2\\
        6 & 6.5 & YS-4 & 1.5 & 35.1$\pm$2.2\\
        7 & 6.5 & YS-6 & 1.0 & 17.2$\pm$0.5\\
        \hline
    \end{tabular}
    \caption{Average light-yields at $HV=V_{break}+3$ V for the configurations under test.}
    \label{lysummary}
\end{table}
\par
\begin{figure}
    \centering
    \includegraphics[width=0.8\linewidth]{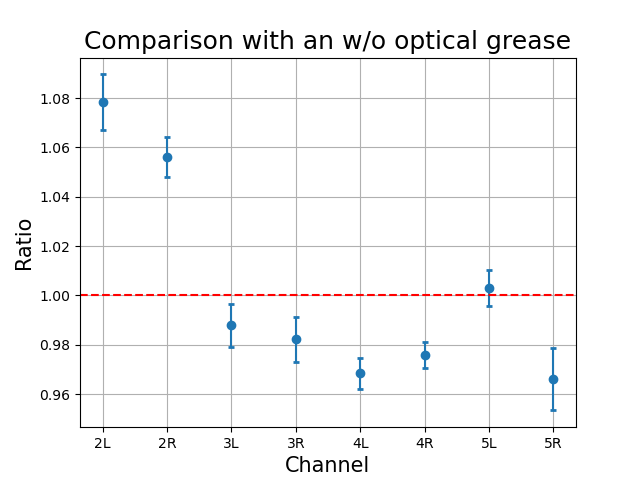}
    \caption{Impact of the optical grease. For eight SiPMs corresponding to four bars, the ratio between the light-yield measured in two different runs is reported. In the second run the two SiPMs of the first bar (corresponding to 2L and 2R) have been matched to the corresponding fibers using optical grease. }
    \label{Grease}
\end{figure}

\noindent The effect of the optical grease in the fiber-SiPM matching has been tested by applying it to the two sides of one bar, and comparing the light-yields with and without grease in the same conditions. The other three bars remained unchanged to provide a control sample. The results are shown in Fig.\ref{Grease}. An improvement in light-yield of the order of 8$\div$10\% is clearly observed.  
\subsection{Attenuation lengths} 
\par
\begin{figure}
    \centering
    \includegraphics[width=1.0\linewidth]{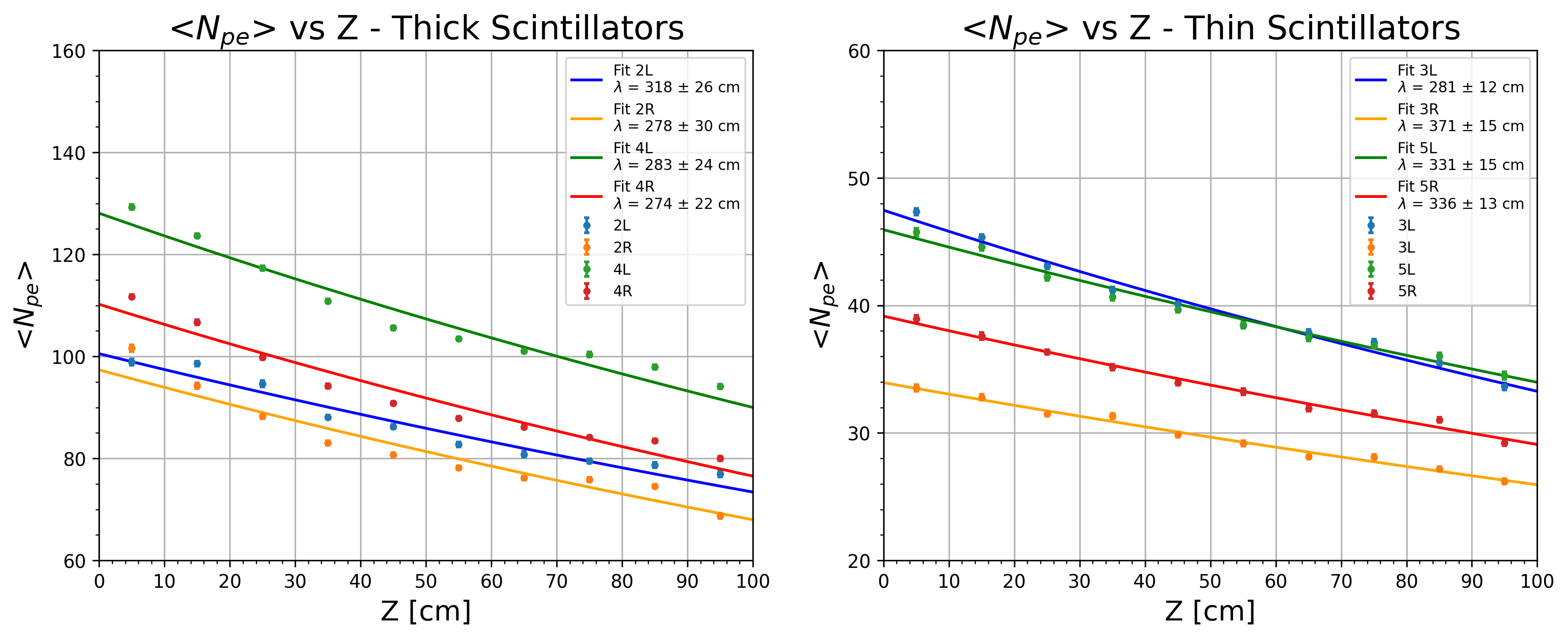}
    \caption{Light-yield as a function of the distance from the SiPM for two Thick bars (configuration 3, left plot) and for two Thin bars (configuration 6, right plot).}
    \label{attlengths}
\end{figure}

\noindent In the light-yield measurements described in the previous section, data are integrated in the full 1 m bar length. Attenuation lengths can be estimated by evaluating the dependence of $N_{pe}$ on the distance $Z$ from the SiPM. For this measurement the bar length is divided in ten sections along $Z$, each 10 cm wide. $N_{pe}$ values are obtained from the amplitude distributions of each section and are plotted as a function of $Z$. Fig.\ref{attlengths} illustrates the observed dependencies for two specific configurations, 3 and 6. Exponential fits are super-imposed. In the case of Thin scintillators, the exponential function describes very well the behavior, and the extracted values of the attenuation lengths are in agreement with those quoted by the fiber manufacturer. The data from the Thick scintillators, on the other hand, show a clear double exponential behavior. Since the bars are instrumented with the same fibers, the different behavior should be attributed to the different scintillators. This effect deserves an additional investigation possibly using longer bars. 

\subsection{Light velocity in the fibers} 
\par
\begin{figure}
    \centering
    \includegraphics[width=0.8\linewidth]{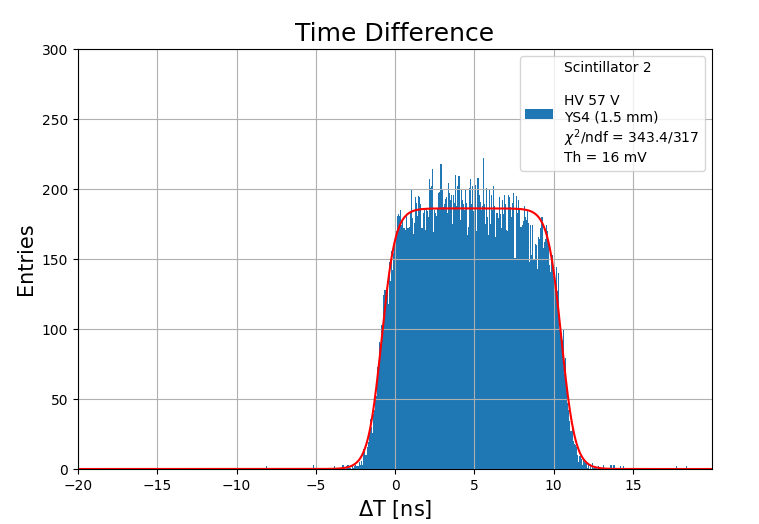}
    \caption{Example of distribution of the time difference $t_L-t_R$ with a double Fermi-Dirac fit superimposed. }
    \label{lightv}
\end{figure}

\noindent Fig.\ref{lightv} shows the distribution of the time differences $t_L-t_R$ for a bar in configuration 3 at the maximum HV of 57 V. The fit superimposed is a double-Fermi-Dirac function. The total width at half height $\Delta t_{width}$ extracted from the fit allows to evaluate the effective light velocity in the fibers: $v_l=2L/\Delta t_{width}$, $L$ being the bar length. The values obtained are in the range between 17.4 and 17.7 cm/ns, consistent among the different fibers used. 
\begin{figure}
    \centering
    \includegraphics[width=0.85\linewidth]{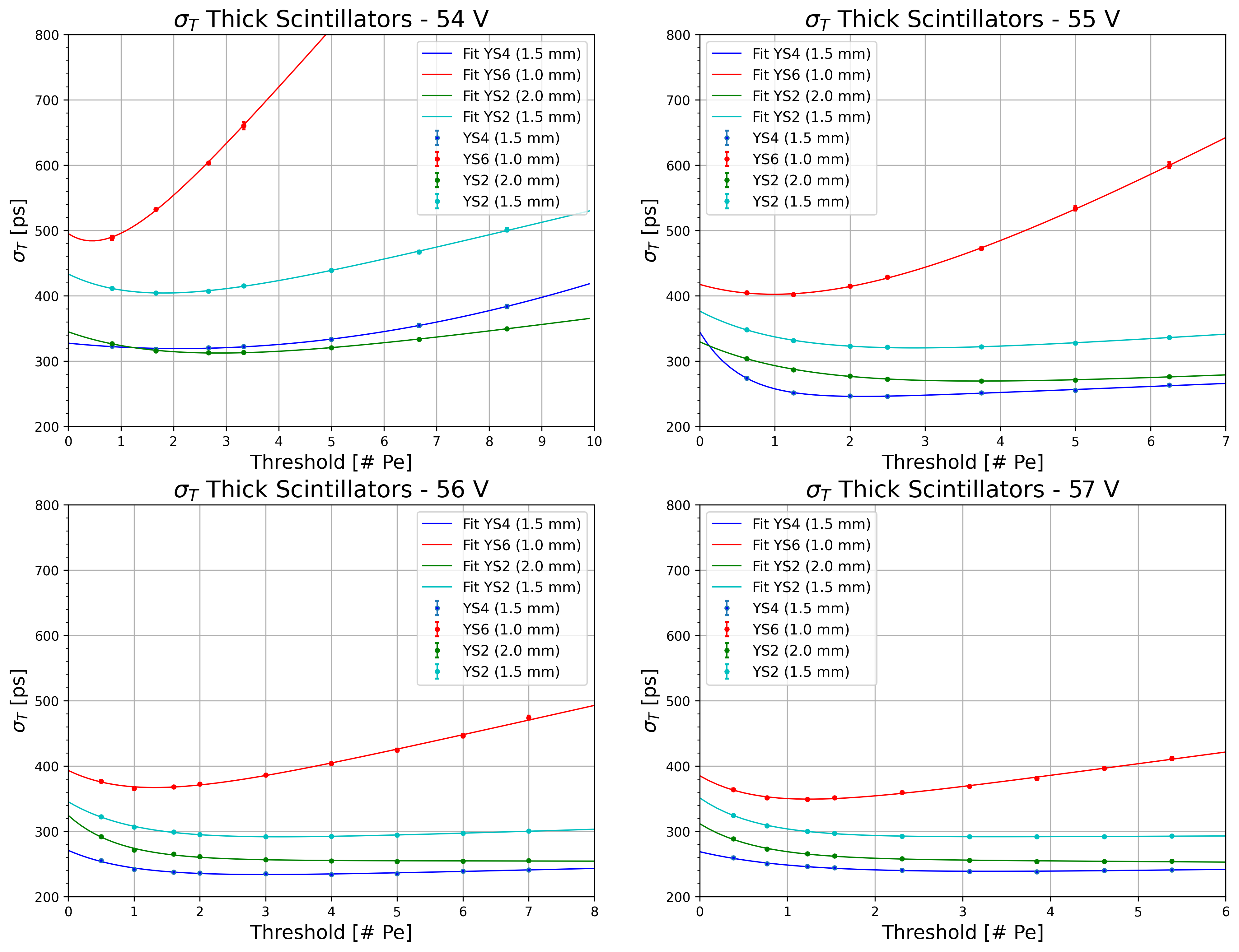}
    \caption{Time resolutions as a function of the threshold in number of photoelectrons for the four configurations, 1 to 4, related to thick scintillators. The four plots correspond to data taken at different HV.}
    \label{sigmatThick}
\end{figure}
\begin{figure}
    \centering
    \includegraphics[width=0.85\linewidth]{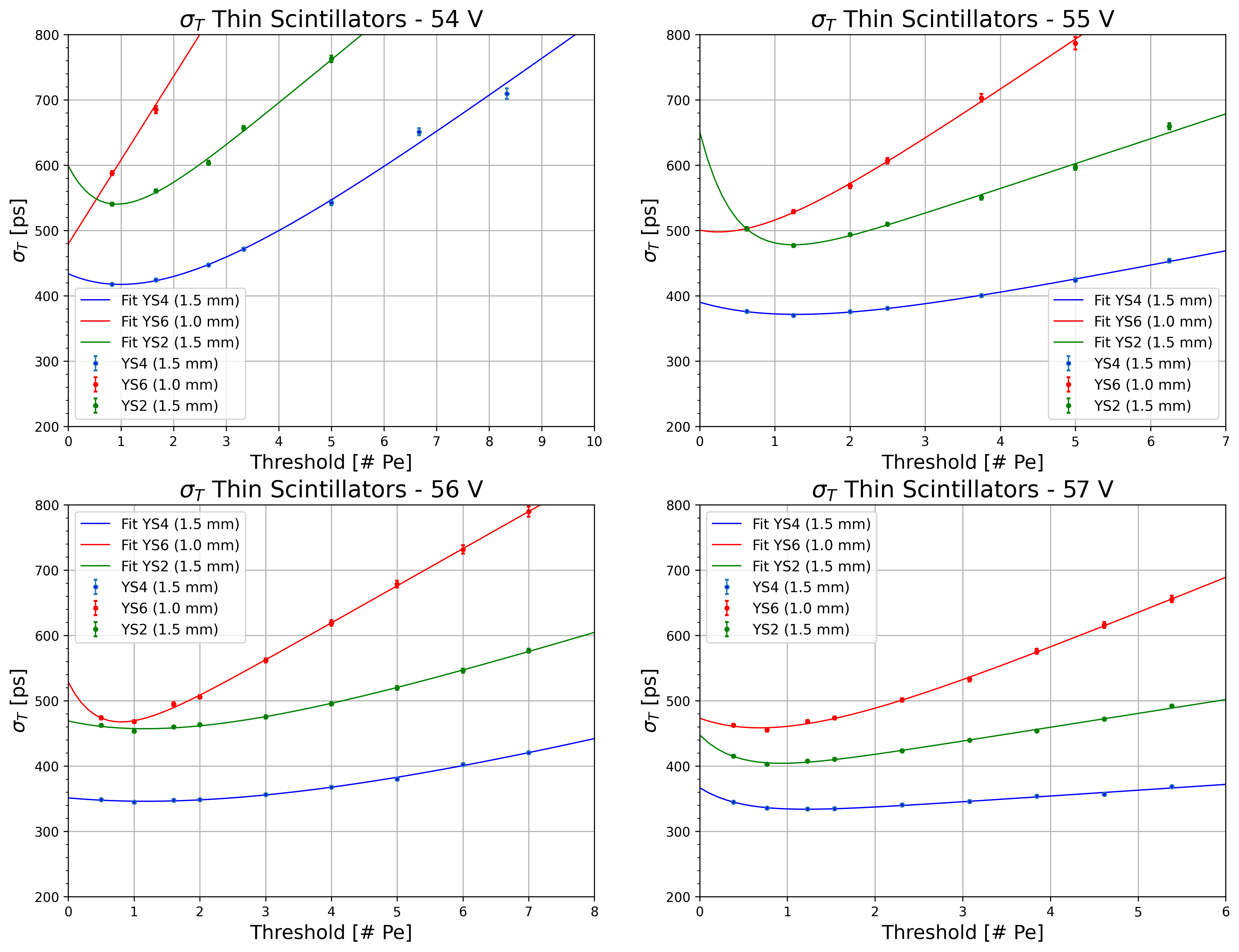}
    \caption{Time resolutions as a function of the threshold in number of photoelectrons for the three configurations, 5 to 7, related to thin scintillators. The four plots correspond to data taken at different HV.}
    \label{sigmatThin}
\end{figure}
\begin{figure}
    \centering
    \includegraphics[width=0.8\linewidth]{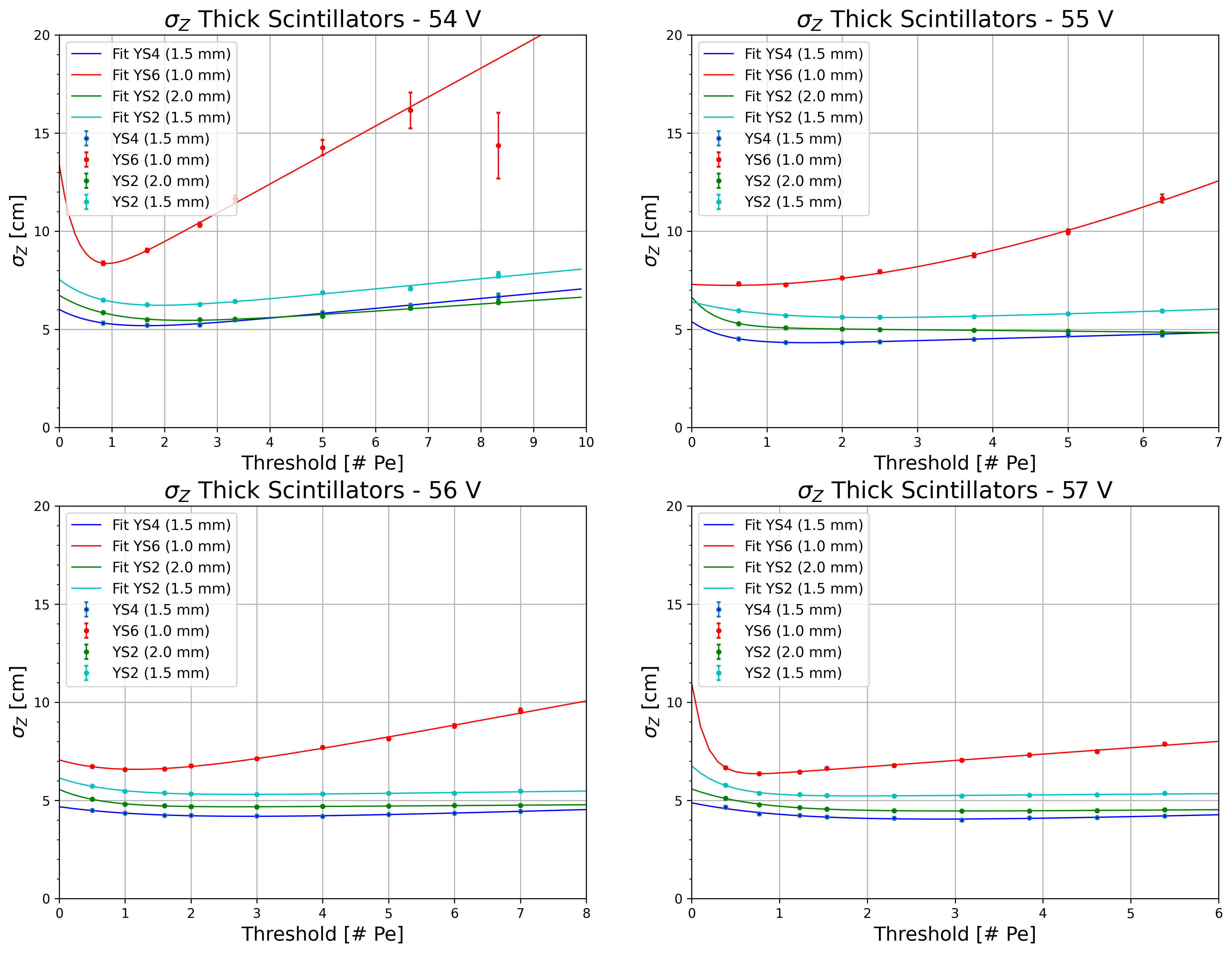}
    \caption{$Z$ resolutions as a function of the threshold in number of photoelectrons for the four configurations, 1 to 4, related to thick scintillators. The four plots correspond to data taken at different HV.}
    \label{sigmaZThick}
\end{figure}
\begin{figure}
    \centering
    \includegraphics[width=0.8\linewidth]{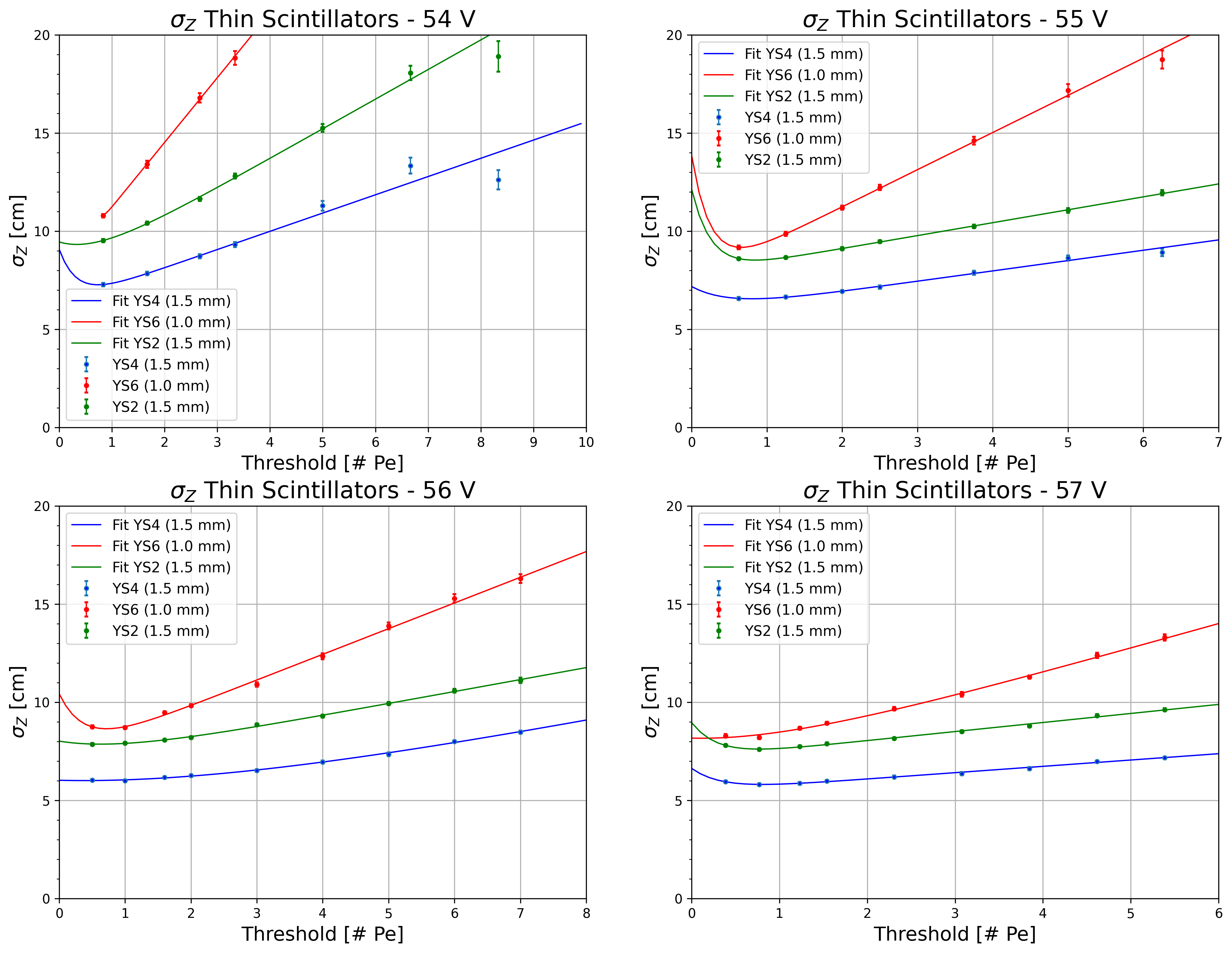}
    \caption{$Z$ resolutions as a function of the threshold in number of photoelectrons for the three configurations, 5 to 7, related to thin scintillators. The four plots correspond to data taken at different HV.}
    \label{sigmaZThin}
\end{figure}
\begin{figure}
    \centering
    \includegraphics[width=0.85\linewidth]{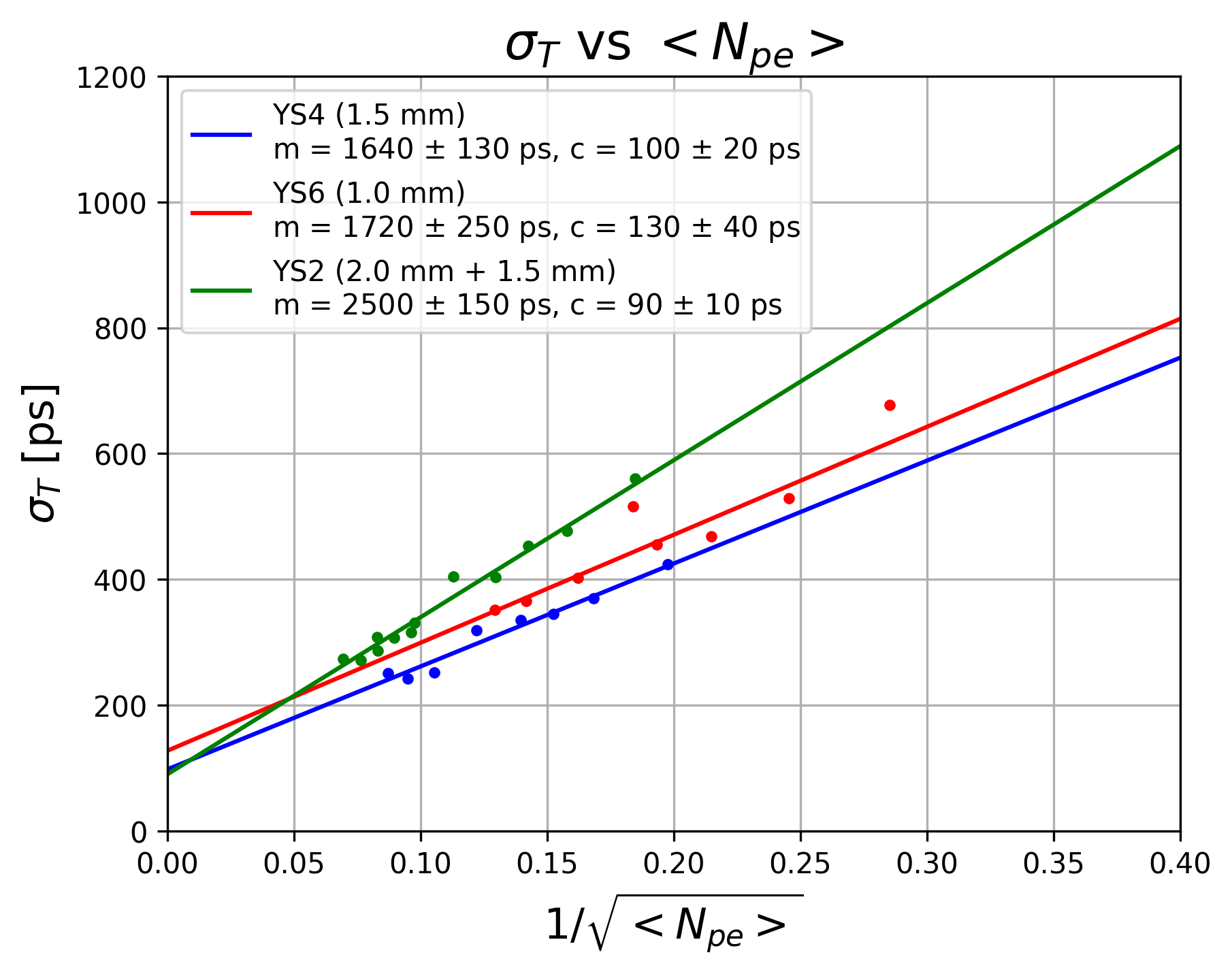}
    \caption{Time resolution as a function of $1/\sqrt{<N_{pe}>}$ for data related to the three kind of fibers. Linear fits are superimposed.}
    \label{scaling}
\end{figure}
\subsection{Time resolution} 
\par\noindent The time resolutions of the different configurations are determined as a function of threshold and HV. The results for all the configurations are summarized in Figs.\ref{sigmatThick} and in Fig.\ref{sigmatThin} for the Thick and Thin bars respectively. The thresholds are given in terms of number of photoelectrons as explained above. For a given configuration the resolution improves with HV due to the PDE increase. The dependence on the threshold is less trivial. In general, resolutions improve by lowering the threshold, but for very small thresholds, around 1-2 photoelectrons, some noise effect can be in place affecting the resolution.

\par\noindent The best resolution is obtained for Thick scintillators instrumented with YS-4 fibers of 1.5 mm diameter. YS-2 fibers of larger diameter give a worse resolution even if they have a larger light-yield. This is due to the larger time constant of the YS-2 fibers with respect to YS-4 (see Table \ref{configurations}). Resolutions at the level of 250 ps can be obtained for these detectors at the baseline voltage (55 V) and for thresholds in the range 1 - 4 photoelectrons.
\par\noindent Thin scintillators allow to reach, in the best configuration, time resolutions around 350 ps, given the lower light-yield.
%    \item 	X resolution (only for runs with triangles)
\par\noindent Similar behaviors are obtained for what concerns the $Z$ resolution. The results are summarized in Figs. \ref{sigmaZThick} and \ref{sigmaZThin} for Thick and Thin scintillators respectively. In the best configurations, resolutions down to around 4 cm can be achieved.

\par\noindent The runs taken with different HV in the same configuration allows to evaluate the scaling of the time resolution with the light-yield. Fig.\ref{scaling} summarizes the dependence of the measured time resolutions $\sigma_t$ on $1/\sqrt{<N_{pe}>}$. In each plot the data from the configurations taken with the same kind of fiber are put together. 
\par\noindent The resolution turns out to scale reasonably well with $1/\sqrt{N_{pe}}$. The slope obtained using the YS-2 fibers are significantly larger than those obtained with YS-4 and YS-6 in agreement with expectations (see Table \ref{configurations}). Constant terms at the level of 100 ps are obtained for all kinds of fibers.

\section{Conclusion}
\par\noindent The design of large area muon systems at next generation $e^+e^-$ colliders, requires to complement detectors with high space resolution for muon momentum measurement with fast detectors providing a time resolution of the order of few hundreds of ps together with a space resolution of the order of 1 mm for the 2nd coordinate and of 1 cm for the orthogonal coordinate.
\par\noindent The study presented in this paper shows that detectors based on a single layer of 1 m long extruded bars equipped with fast WLS fibers readout through SiPMs allow to reach time resolutions down to 250 ps and resolutions on the longitudinal coordinate down to 4 cm on the full area. The dependence of the resolutions on the HV of the SiPMs and on the threshold applied to the signals is shown in detail.
\par\noindent The light-yield turns out to scale with the scintillator thickness and with the fiber diameter, and depends, even if to a minor extent, on the optical contact between the fiber and the SiPM. 
\par\noindent The measurement of the transverse coordinate with the required resolutions requires either different geometries like the one proposed in \cite{Allegro}, or multi-layer structures. 
\section*{Acknowledgements}
\par\noindent We gratefully acknowledge the support of INFN for this research. We thank the Mechanical Workshop and the Mechanical Design Laboratory of the INFN Sezione di Roma for designing and preparing the SiPM supports and the light-tight box. We thank the Electronics Laboratory of the INFN Sezione di Roma for designing and producing the Single Channel Amplifier Boards.


\newpage

\begin{thebibliography}{10}
\bibitem{FCCee} A. Abada et al., " FCC-ee: The Lepton Collider: Future Circular Collider Conceptual Design Report" Volume 2, Eur. Phys. J. ST 228, 261 (2019);
\bibitem{briefing} R.Forty et al. (Eds) "Physics Briefing Book", CERN-2025-008, CERN-ESU-2025-001;
\bibitem{Allegro} F.Anulli et al., "A High-Precision, Fast, Robust, and Cost-Effective Muon Detector Concept for the FCC-ee " arXiv:2504.10448;
\bibitem{Pyramid} L.Aliaga et al., "Design, Performance and Calibration of the MINERvA detector", Nucl. Instr. and Meth. A743, (2014), 130-159;
\bibitem{Belle2} T.Aushev et al., "A scintillator based $K_L$ and muon detector for Belle II experiment", Nucl. Instr. and Meth. A789, 11 (2015), 134-142;
\bibitem{Russi1} D.Denisov, V.Evdokimov, S.Lukić, " Time and position resolution of the scintillator strips for a muon
system at future colliders", Nucl. Instr. and Meth. A823, (2016), 120-125;
\bibitem{Russi2} D.Denisov, V.Evdokimov, S.Lukić, P.Ujic, "Test-beam studies of the light-yield, time and coordinate resolutions of scintillator strips with WLS fibers and SiPM readout", Nucl. Instr. and Meth. A848, (2017), 54-59;
\bibitem{Mu2e} Artikov et al., " Photoelectron yields of scintillation counters with embedded wavelength-shifting fibers read out with silicon photomultipliers" Nucl. Instr. and Meth. A890, (2018), 84-95;
\bibitem{Bross} A.D.Bross et al., "Tomographic Muon Imaging of the Great Pyramid of Giza", JAIS-280 (2022);
\bibitem{DRS} Full custom integrated circuit developed at PSI, Switzerland, see S.Ritt et al., "Application of the DRS chip for fast waveform digitizing" Nucl.Instr. and Meth. A623 (2010) 486-488;
\bibitem{Fermilab} A. Pla-Dalmau, A.D. Bross, V. Rykalin, "Extruding plastic scintillator at fermilab", IEEE NSS Conference Record (2003), FERMILAB-CONF-03-318-E;
\bibitem{japanese} S.Kodoma et al, " Performance of new Kuraray
wavelength-shifting fibers with short decay time", Prog. Theor. Exp. Phys. 2023. 
\end{thebibliography}
\end{document}